\documentclass[journal]{IEEEtran} 				
\IEEEoverridecommandlockouts 						
\usepackage{amsmath,amssymb}
\usepackage{amsthm}
\usepackage{color}
\usepackage[mathscr]{eucal}
\usepackage{graphics,graphicx,multicol}
\usepackage{epsfig}
\usepackage{enumerate}
\usepackage{subfigure}
\usepackage{algorithmic}
\usepackage{algorithm}
\usepackage{morefloats}
\usepackage{enumerate}
\usepackage{subfigure}
\usepackage{multirow}
\usepackage{array}
\usepackage{pstricks, pst-node, pst-plot, pst-circ}
\usepackage{moredefs}
\usepackage{tabu}
\usepackage{flexisym}

\usepackage{courier}

\begin{document}

\title{Optimal Power Allocation for LTE Users with Different Modulations}
 \author{\IEEEauthorblockN{Ying Wang\IEEEauthorrefmark{1}, Ahmed Abdelhadi\IEEEauthorrefmark{2}}\\
     \IEEEauthorblockA{\IEEEauthorrefmark{1}EECS, University of Michigan, wying@umich.edu},\\
     \IEEEauthorblockA{\IEEEauthorrefmark{2}Hume Center, Virginia Tech, aabdelhadi@vt.edu}
 }
\maketitle

\begin{abstract}
In this paper, we demonstrate the optimal power allocation for QPSK, 16-QAM, and 64-QAM modulation schedules and the role of channel quality indicator (CQI). We used sigmoidal-like utility functions to represent the probability of successful reception of packets at user equipment (UE). CQI as a feedback to the base station (BS) indicates the data rate that a downlink channel can support. With Levenberg-Marquardt (LM) Optimization method, we present utility functions of different CQI values for standardized 15 Modulation order and Coding Scheme (MCS) in $3^{rd}$ Generation Partnership Project (3GPP). Finally, we simulate and show the results of the optimal power allocation algorithm.
\end{abstract}

\begin{keywords}
Convex Optimization, Resource Allocation, CQI, SNR, Parameter Identification
\end{keywords}
\pagenumbering{gobble}

\section{Introduction}\label{sec:intro}

The research area of optimizing the resource allocation in cellular networks has received significant attention. Due to an increasing need for wireless adaptive real-time applications, the current and merging standards are supporting various higher modulation schemes. For example, long term evolution (LTE), the fourth-generation (4G) wireless standard specified by 3rd Generation Partnership Project (3GPP), supports higher modulation schemes such as QPSK, 16-QAM, and 64-QAM according to LTE-Advanced (LTE-A). The solution for an optimal and fair resource allocation is demanded by users to seek better quality of services (QoS), where QoS can be the minimum successful transmission probability. For this reason, numerous research efforts have been made to optimally allocate power in order to achieve a better certain signal-to-noise ratio (SNR), and guarantee minimum successful transmission probability of packets.

The network utility maximization framework can be used to improve power allocation and achieve better QoS.
The utility function is the probability of successful reception of packets versus power. It is considered as a controlling parameter through which a user's QoS can be guaranteed. And it is a representation of the QoS for a user. The goal of the network utility maximization framework is, therefore, to allocate power in order to maximize network utility, which is defined as a product of all users' utilities. In \cite{lee2005downlink}, the authors defined utility functions to maximize signal-to-interference-plus-noise ratio (SINR), and it is represented by the sigmoidal-like function that proposed in \cite{shenker1995fundamental}.

It is difficult to design resource allocation algorithms that maximize system efficiency, ensure fairness, and meet the QoS requirements of all users because of the randomness in the wireless broadcast channel. Research work, for example in \cite{viswanath2002opportunistic}, used opportunistic resource allocation algorithms to improve the system efficiency, but they failed to address the QoS requirements of users and maintain a fair allocation of resources among users.

The Channel Quality Indicator (CQI) is a feedback from UE the BS and it indicates the data rate that can be supported by the downlink channel. Thus, it can be used to deliver different utility functions for different CQIs and, therefore, simulate the optimal power allocation in the cellular network.

The main contributions of this paper are 1) we mapped the standard LTE CQI values to utility functions and the corresponding distances from the BS, 2) we used LM optimization method to identify the parameters of the utility functions and 3) we simulated the optimal power allocation among 15 UEs and each of them has a different CQI in the cellular network.

\subsection{Related Work}\label{sec:related}
The authors in \cite{JWlee} developed a utility-based optimal downlink power allocation algorithm for multi-class wireless networks. Moreover, they used a numerical approach to show that its performance is close to that of the global optimal power allocation. However in theses two studies the fairness among the users is not considered, e.g. the minimal quality of service (QoS) is not guaranteed for all users. 

In \cite{Jang-Won}-[6], the fairness in resource allocation is considered as a more important issue. The fairness/QoS constraints have been set in the optimal resource allocation in \cite{Jang-Won} and \cite{Liu}. In \cite{Jang-Won}, the authors provided an opportunistic power scheduling scheme for "multi-server" wireless systems while meeting the minimum QoS for each user. A stochastic process has been used to present each user's performance value in \cite{Liu2} and they proposed an opportunistic transmission-scheduling policy to maximize the average system performance. In \cite{liu2003opportunistic}, Multi-channel Fair Scheduler (MFS) has been introduced and analyzed to guarantees both long-term deterministic (MFS-D) and probabilistic (MFSP) fairness over multiple wireless channels. They provided a framework that maximizes total system throughput for opportunistic scheduling over multiple wireless channels.

Later in  \cite{abdelhadi2014optimal}, the study introduced a novel approach for power allocation in the cellular network where the user's utility function has been modeled as sigmoidal-like function. In this work, the power allocation optimization problem is formulated as a product of the utilities of all users with utility proportional fairness policy. A priority has been assigned to users with lower modulation schemes, at the same time giving non-zero power allocation to users using higher modulation schemes. A similar method was used to allocate optimal rates in \cite{Haya1}, \cite{Haya2}

The MCS selection has been widely studied. Work in \cite{goldsmith1998adaptive}-\cite{kim2008optimal} has proposed adaptive modulation and coding (AMC) to enhance the system throughput according to the channel quality. The CQI, the only feedback to BS, corresponds to a resource block (RB) or multiple RB's in the form of MCS index \cite{fan2011mcs}. And CQI value provides important information in link adaption. The study in \cite{fan2011mcs} developed several MCS selection schemes for downlink transmission in LTE systems by using the effective packed-level SINR. Thresholds were set to the SINR values with the Block Error Rate (BLER) smaller than 10\% for the MCSs and mapping between SINR value, and CQI were also provided. In \cite{ibrahimchannel}, CQI values were derived by calculating the post-detection SINR from the instantaneous channel quality measured at the receiver side.  

This paper is organized as follows. We first introduce the system model set up in Section \ref{sec:SM}. Then we review the CQI with more details in Section \ref{sec:CQI}. In Section \ref{sec:Mapping}, we describe the solution for mapping the CQI values to the utility function. Section \ref{sec:opa} that briefly describes the optimal power allocation algorithm that was proposed in \cite{abdelhadi2014optimal}. In Section \ref{sec:sim}, we discuss the simulation set up and the results along with a discussion. Finally, Section \ref{sec:conclude} concludes this paper.

\section{System Model}\label{sec:SM}
In this paper, we consider a single cellular system consisting of a single BS and $M$ UEs. Each UE has a CQI based on its location in the cell, where the UE closer to the BS has a higher CQI. As shown in Table \ref{tab:title}, higher CQI corresponds to higher modulation. Our goal is to generate different utility functions with respect to CQI values to represent the QoS of users and optimally allocate powers to UEs. The total power at BS is $P_{T}$.

\section{Channel Quality Indicator (CQI)}\label{sec:CQI}

CQI is the feedback that the UE sends to indicate the data rate which can be supported by the downlink channel. The BS selects an appropriate modulation scheme and code rate for downlink transmission based on CQI values.

\begin{table}[!t]
\renewcommand{\arraystretch}{1.3}
\caption{CQI Overview} \label{tab:title} 
\centering
\begin{tabular}{ | l | l | l | l | }
\hline
CQI Index & Modulation & Code Rate X 1024 & Efficiency \\ \hline
0 & No transmission &  &  \\ \hline
1 & QPSK & 78 & 0.1523 \\ \hline
2 & QPSK & 120 & 0.2344 \\ \hline
3 & QPSK & 193 & 0.3880 \\ \hline
4 & QPSK & 308 & 0.6016 \\ \hline
5 & QPSK & 449 & 0.8770 \\ \hline
6 & QPSK & 602 & 1.1758 \\ \hline
7 & 16QAM & 378 & 1.4766 \\ \hline
8 & 16QAM & 490 & 1.9141 \\ \hline
9 & 16QAM & 616 & 2.4063 \\ \hline
10 & 64QAM & 466 & 2.7305 \\ \hline
11 & 64QAM & 567 & 3.3223 \\ \hline
12 & 64QAM & 666 & 3.9023 \\ \hline
13 & 64QAM & 722 & 4.5234 \\ \hline
14 & 64QAM & 873 & 5.1152 \\ \hline
15 & 64QAM & 948 & 5.5547 \\ 
\hline
\end {tabular}
\end{table}

In addition to indicating the downlink channel quality, CQI also accounts for the capabilities of the UE's receiver. Table \ref{tab:title} \cite{CQI_1}, \cite{CQI_2} shows the corresponding modulation scheme, code rate and efficiency for different CQI values.

\section{Mapping CQI to Utility Function}\label{sec:Mapping}
This section provides the method that we used to map the CQI values to the utility function. 

\subsection{SNR Probability}
\begin{figure}[!t]
\centering
\includegraphics[width=3.5in]{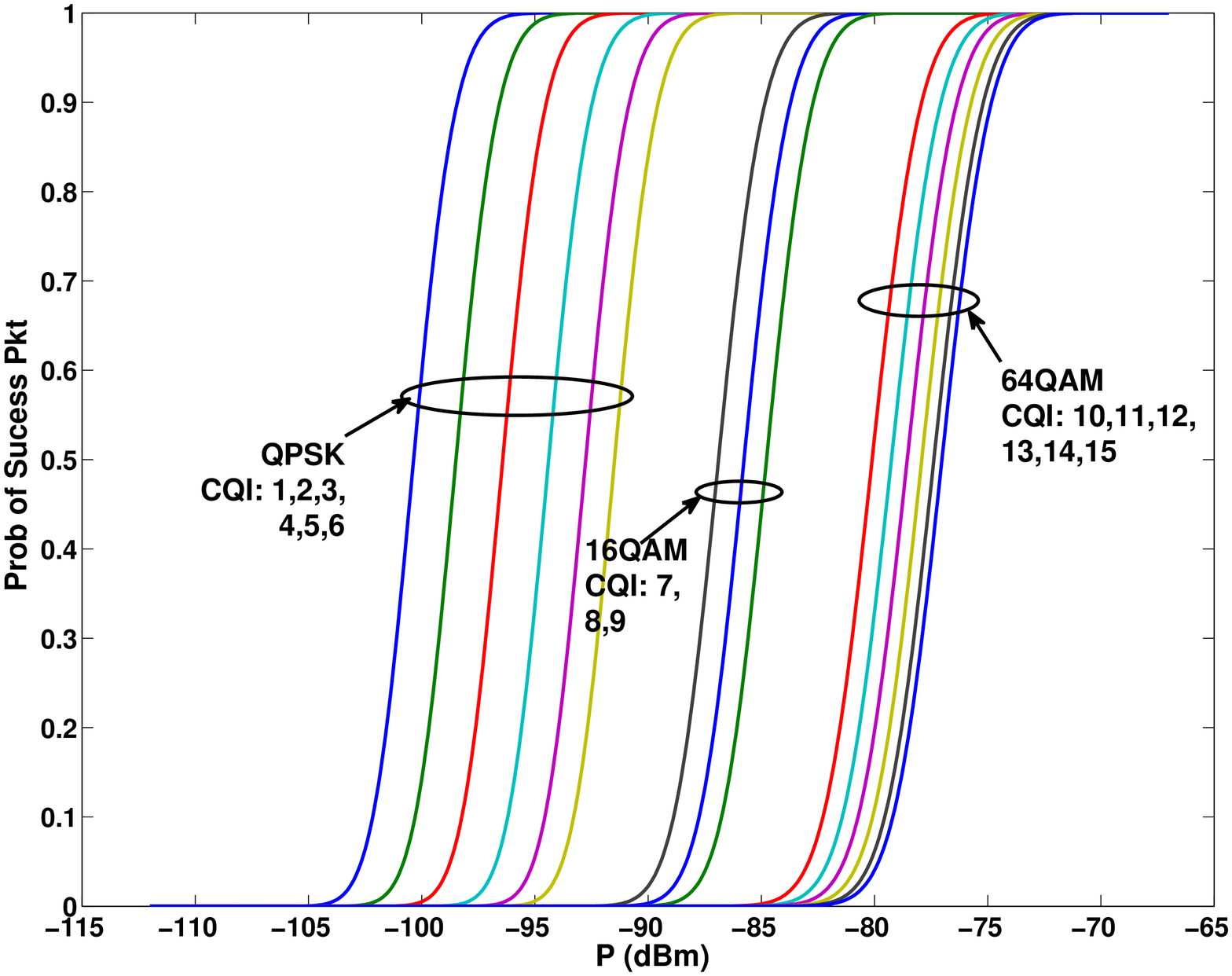}
\caption{Cumulative distribution function of successful packet transmission for different modulation schemes with different CQIs.}
\label{fig_SNR_PRO}
\end{figure}

With the efficiency values that correspond to different CQIs in Table \ref{tab:title}, the probability of successful receiving packages for different SNR values are calculated. The probability of successful packages transmission as a function of the receiver power (dBm) is obtained in the Figure \ref{fig_SNR_PRO}, where the receiver power (dBm) is obtained by adding the downlink receiver noise floor to the SNR.

\subsection{Mapping CQI SNR with distance}
In \cite{CQILINEAREQN}, a linear function (\ref{eqn:SNR}) has been proposed to map SNR to CQI. Then after flooring, the CQIs, obtained by the linear function (\ref{eqn:SNR}), over all resource blocks (RB) are reported back to the BS.

\begin{equation}\label{eqn:SNR}
\text{CQI} = 0.5223 \text{SNR} +4.6176
\end{equation}

\subsection{Transmitter Power Utility function}
Path-loss, known as the power reduction through space, limits the power that a recipient can receive at a distance from the BS. The further the UE is, the lower power it can receive. UE power is limited by (\ref{eqn:Pathloss}), where $f$ is the carrier frequency and $c$ is speed of the light. In urban environment $\alpha$ is 3.5. The UE power drops as distance $d$ gets larger. 

\begin{equation}\label{eqn:Pathloss}
P_{\text{UE}} = \frac{P_{\text{BS}} f}{c (4 \pi d)^\alpha} 
\end{equation}

Our optimal power allocation method is to find the optimal power at BS that can be distributed to the users. Therefore, a conversion is needed to convert the receiver power to the transmitter power. This conversion is done using Equation (\ref{eqn:Pathloss}).

\subsection{Parameterization Utility Function}\label{sec: PUF}
In our model, we use the normalized sigmoidal-like utility functions, as in \cite{DBLP:journals/corr/Abdel-HadiC14a} and \cite{DBLP:journals/corr/ShajaiahAC14}, it can be expressed as 
\begin{equation}\label{eqn:sigmoid}
U_i(P_i) = c_i\Big(\frac{1}{1+e^{-a_i(P_i-b_i)}}-d_i\Big)
\end{equation}
where $c_i = \frac{1+e^{a_ib_i}}{e^{a_ib_i}}$ and $d_i = \frac{1}{1+e^{a_ib_i}}$.\\

We use the Levenberg-Marquardt (LM) Optimization method to identify the parameters in the utility functions for different CQIs. The LM algorithm uses the approximated Hessian and the information in the gradient, taking into account some regularization factors. We form the error function $E$ as

\begin{equation}\label{eqn:Error_LM}
E(\mathbf{p}) = \sum_{i=1}^{n}[U_i - f(P_i,\mathbf{p})]^{2}
\end{equation}
where $\textbf{p}=[a,b]^T$ and $n$ is the number of data points.\\

The object is to minimize the error function $E$ with respect to the vector $\textbf{p}$. The optimal solution for $\textbf{p}$ is searched by iterations. The updated equation in LM is
\begin{equation}\label{eqn:p_iter}
[J_{f}(\mathbf{p}_{i})^{T}J_{f}(\mathbf{p}_{i} + \lambda I] = 
 -J_{f}(\mathbf{p}_{i})^{T}f(P_i,\mathbf{p_i})
\end{equation}
where $\lambda$ is the damping parameter. 

At the beginning of the iterations, a large value is set to $\lambda$, in this way the robustness is manifested therefore the initial guess can by chosen with less caution. For each iteration, if $E(\mathbf{p_i}+\Delta \mathbf{p_i}) < $ $E(\mathbf{p_{i-1}}+\Delta \mathbf{p_{i-1}})$,  it speeds up the convergence by decreasing $\lambda$ to a certain amount, otherwise, it increases $\lambda$ to enlarge the trust region \cite{LM}.

\section{Optimal Power Allocation}\label{sec:opa}
In \cite{abdelhadi2014optimal}, the optimal utility proportional fairness power allocation problem is formulated as  
\begin{equation}\label{eqn:opt}
\begin{aligned}
& \underset{\textbf{P}}{\text{max}}
& & \prod_{i=1}^{M}\log(U_i(\gamma_i(P_i))) \\
& \text{subject to}
& & \sum_{i=1}^{M}P_i \leq P_T\\
& & &  P_i \geq 0, \;\;\;\;\; \text{for} \; \; i = 1,2, ...,M \; \; \text{and} \; \; P_T \geq 0.
\end{aligned}
\end{equation}
where $P_T$ is the total power of the BS, $M$ is the number of UEs and $\textbf{P} = \{P_1, P_2, ..., P_M\}$.

The utility functions $\log(U_i(\gamma_i(P_i)))$ in the optimization problem (\ref{eqn:opt}) are strictly concave functions and therefore there exists a unique tractable global optimal solution. We used the optimal power allocation algorithm in \cite{abdelhadi2014optimal} and this algorithm is divided into an UE algorithm shown in Algorithm (\ref{alg:UE_FK}) and an BS algorithm shown in Algorithm (\ref{alg:eNodeB_FK}). In this algorithm each user starts with an initial bid $w_i(1)$ and transmits it to BS. The BS initials $w_i(0) = 0$ and compares the difference between the current bid $w_i(n)$ and the previous bid $w_i(n-1)$ with a threshold $\delta$. If it is smaller than $\delta$, then it exits the process and allocates the optimal power $P_{i}^{\text{opt}}=\frac{w_i(n)}{p(n)}$ to UE. Otherwise, the BS calculates the shadow price $p(n) = \frac{\sum_{i=1}^{M}w_i(n)}{R}$ and sends it back to the UEs. And after each UE receives the shadow price, it calculates the power $P_{i}$ that maximizes $(log U_{i}(\gamma_{i}(P_{i}))-p(n)P_{i})$ and generates new bids $w_i(n) = p(n)P_{i}(n)$. After that, each UE compares $|w_i(n)-w_i(n-1)|$ to a pre-set threshold $\big\{\Delta w(n) = l_1 e^{n/l_2} \big \}$. If the value is greater than the threshold, then the UE recalculates a bit $w_{i,new}(n) = w_i(n-1)+\text{sign}(w_i(n)-w_i (n-1)) \Delta w(n)$ and sends this new value to the BS, otherwise it will send $w_i(n)$ to the BS. This process repeats until $|w_i(n) -w_i(n-1)|<\delta$.


\begin{algorithm}
\caption{UE Algorithm \cite{abdelhadi2014optimal}}\label{alg:UE_FK}
\begin{algorithmic}
\STATE {Send initial bid $w_i(1)$ to BS}
\LOOP
  \STATE {Receive shadow price $p(n)$ from BS}
  \IF {STOP from BS} %
    \STATE {Calculate allocated rate $P_i ^{\text{opt}}=\frac{w_i(n)}{p(n)}$}
    \STATE {STOP}
  \ELSE
    \STATE {Solve $P_{i}(n) = \arg \underset{{P_{i}}}{\text{max}}(\log U_{i}(\gamma_{i}(P_{i}))-p(n)P_{i})$}
    \STATE {Calculate new bid $w_i(n) = p(n)P_{i}(n)$}
    \IF {$|w_i(n)-w_i(n-1)| >  \Delta w(n)$}  %
    \STATE{$w_i(n) = w_i(n-1)+\text{sign}(w_i(n)-w_i (n-1)) \Delta w(n)$}
    \STATE{$\big\{\Delta w(n) = l_1 e^{n/l_2} \big \}$}
     \ENDIF 
    \STATE {Send new bid $w_i (n)$ to BS}
  \ENDIF 
\ENDLOOP
\end{algorithmic}
\end{algorithm}


\begin{algorithm}
\caption{BS Algorithm \cite{abdelhadi2014optimal}}\label{alg:eNodeB_FK}
\begin{algorithmic}
\LOOP
  \STATE {Receive bids $w_i(n)$ from UEs}
  \COMMENT{Let $w_i(0) = 0\:\:\forall i$}
      \IF {$|w_i(n) -w_i(n-1)|<\delta  \:\:\forall i$} %
        \STATE {Allocate rates, $P_{i}^{\text{opt}}=\frac{w_i(n)}{p(n)}$ to user $i$}  
        \STATE {STOP} 
    \ELSE
  \STATE {Calculate $p(n) = \frac{\sum_{i=1}^{M}w_i(n)}{R}$}
  \STATE {Send new shadow price $p(n)$ to all UEs}
  \ENDIF 
\ENDLOOP
\end{algorithmic}
\end{algorithm}

\section{Simulation Results}\label{sec:sim}
\begin{figure}[!t]
\centering
\includegraphics[width=3.5in]{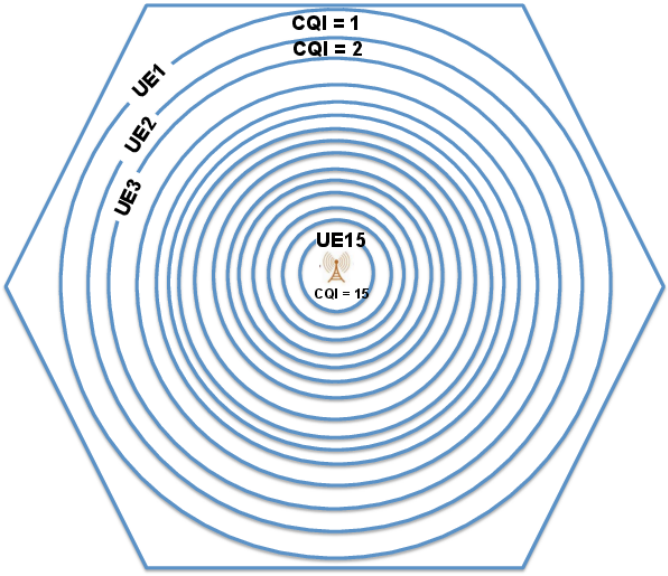}
\caption{Simulation setup}
\label{fig_simu}
\end{figure}

We simulated the cell network with 1 BS and $M$ UEs where $M = 15$. The total power $P_{T}$ at the BS is 40W which is a typical value for macro cell base station at the antenna connector. 15 UEs are placed at different locations in the cell and one UE in each CQI zone, as seen in Figure \ref{fig_simu}. The UEs are placed at the further edge of each CQI zone, which is the worst channel quality in each CQI zone.

\begin{figure}[!t]
\centering
\includegraphics[width=3.5in]{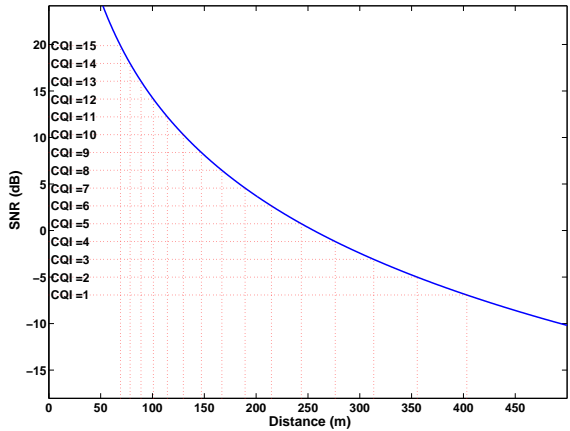}
\caption{Mapping CQI, SNR with distance to the BS}
\label{fig_CQI_DIST}
\end{figure}

The mapping among CQI, SNR and distance is plotted in Figure \ref{fig_CQI_DIST}. As the UE moves further away the CQI decreases and so as SNR. The UEs, standing in the range between 1m and 69.14m away from the BS, have the best channel quality (CQI = 15) with 64-QAM. Whereas the CQI zone for the worst channel quality (CQI = 1) is 355.5m to 403.2m away from the BS.

\begin{figure}[!t]
\centering
\includegraphics[width=3.5in]{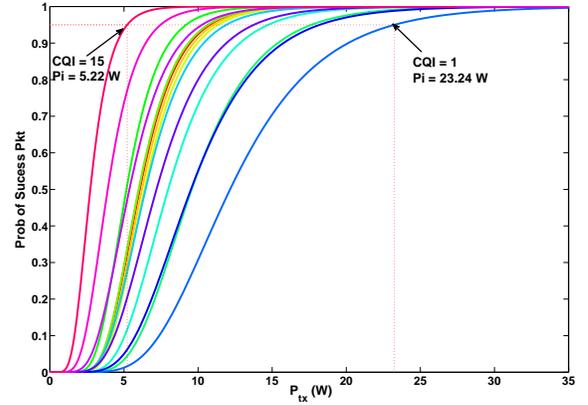}
\caption{Power Utility Function}
\label{fig_power_utility}
\end{figure}

\begin{figure}[!t]
\centering
\includegraphics[width=3.5in]{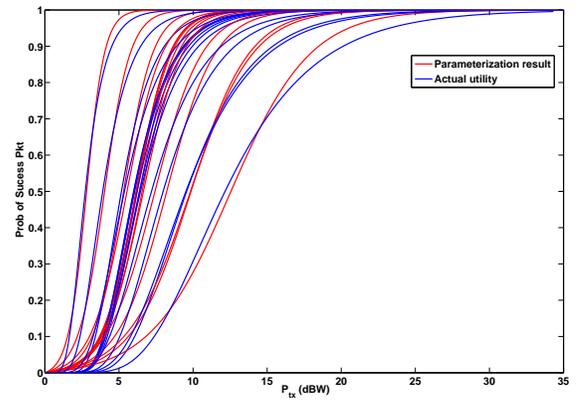}
\caption{Parameterization Power Utility Function}
\label{fig_para}
\end{figure}

The probabilities of the successful package, known as the utility functions, for different CQIs, are shown in Figure \ref{fig_power_utility}. It is a function of the transmitter power. It looks like the sigmoidal-like function, therefore we can use the method in Section \ref{sec:opa}. And as shown in Figure \ref{fig_power_utility}, we set the minimum QoS requirement to be achieving at least a 95\% successful packet transmission. For UE with CQI 15, the power that required to achieve the minimum QoS is about 5.22W whereas the UE with CQI 1 needs 23.24W to have the minimum QoS. To curve fit using the LM algorithm in section \ref{sec: PUF}, the results are shown in Figure \ref{fig_para} and the parameters of 15 utility functions are displayed in Table \ref{tab:title1}. The mean square error (MSE)s of the parameterization are also shown in Table \ref{tab:title1}.

\begin{table}[!t]
\renewcommand{\arraystretch}{1.3}
\caption{Utility parameters} \label{tab:title1} 
\centering
\begin{tabular}{ | l | l | l | l | l |}
\hline
CQI Index & Modulation & a & b & MSE\\\hline
1 & QPSK & 0.8676 & 6.2257 & 4.2188E-4\\ \hline
2 & QPSK & 0.8761 & 6.1657 & 3.8427E-4\\ \hline
3 & QPSK & 0.8466 & 6.3812 & 3.5274E-4\\ \hline
4 & QPSK & 0.8244 & 6.5526 & 3.2596E-4\\ \hline
5 & QPSK & 0.8789 & 6.1467 & 3.0182E-4\\ \hline
6 & QPSK & 1.0188 & 5.3029 & 2.8198E-4\\ \hline
7 & 16QAM & 0.5077 & 9.8303 & 2.8698E-4\\ \hline
8 & 16QAM & 0.6086 & 8.1999 & 2.7031E-4\\ \hline
9 & 16QAM & 0.7524 & 6.6333 & 2.5546E-4\\ \hline
10 & 64QAM & 0.3697 & 12.5005 & 2.5862E-4\\ \hline
11 & 64QAM & 0.4722 & 9.7873 & 2.4527E-4\\ \hline
12 & 64QAM & 0.6248 & 7.3974 & 2.3374E-4\\ \hline
13 & 64QAM & 0.8376 & 5.5177 & 2.2324E-4\\ \hline
14 & 64QAM & 1.1510 & 4.0153 & 2.1364E-4\\ \hline
15 & 64QAM & 1.6471 & 2.8058 & 2.0938E-4\\
\hline
\end {tabular}
\end{table}

\begin{figure}[!t]
\centering
\includegraphics[width=3.5in]{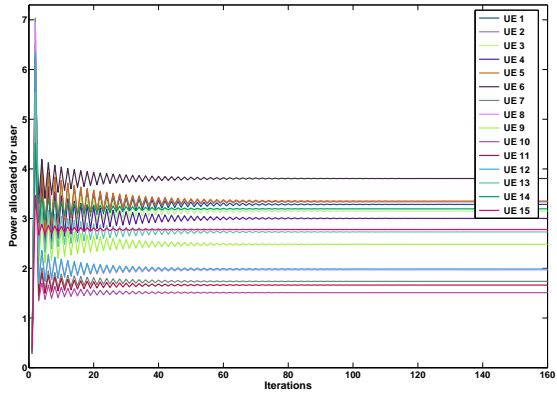}
\caption{Power allocation to 15 different CQI users from BS power 40W}
\label{fig_PA_power}
\end{figure}

\begin{figure}[!t]
\centering
\includegraphics[width=3.5in]{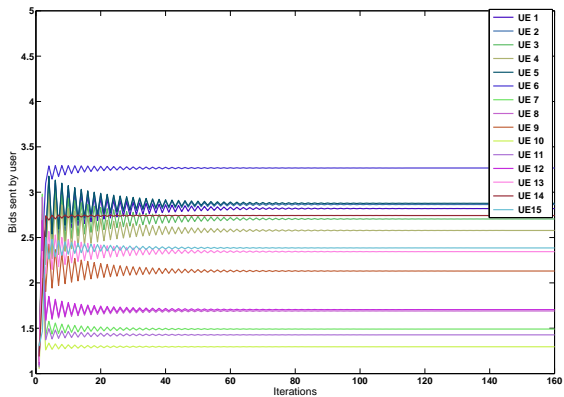}
\caption{Bids sent by 15 different CQI users to BS power 40W}
\label{fig_PA_bids}
\end{figure}

Each UE sends out an initial bid of 10W to the BS, and after applying the optimal power allocation algorithm \cite{abdelhadi2014optimal}, each user is able to receive the optimized power after sending several bids to the BS. The optimal power allocation is plotted in Figure \ref{fig_PA_power}. And in Figure \ref{fig_PA_bids} the iterations of the bidding process are plotted for all UEs.

\section{Conclusion}\label{sec:conclude}
In this paper, we mapped the standard 3GPP CQI values to the sigmoidal-like utility functions, and found the parameters that minimize the MSE. We applied the optimal power allocation algorithm \cite{abdelhadi2014optimal} for realistic cases. We also found that the user with better channel quality would require less power to receive the minimum QoS. In addition, we mapped the CQI and SNR with the distance from the BS. Finally, we demonstrated that by using the optimal power allocation algorithm \cite{abdelhadi2014optimal}, each user was allocated an optimal power.

\bibliographystyle{ieeetr}
\bibliography{paper1cit}
\end{document}